\documentclass[aps,pra,reprint,amsmath,amssymb,superscriptaddress,nobibnotes,showpacs]{revtex4-1}

\usepackage{graphicx}
\usepackage{amsmath}
\usepackage{gensymb}
\usepackage{upgreek}
\usepackage{tabularx}
\usepackage{tabularx,color}
\usepackage{multirow}
\usepackage{hyperref}

\makeatletter

\newcommand{\Rmnum}[1]{\expandafter\@slowromancap\romannumeral #1@}
\makeatother

\begin{document}

\title{Plug-and-Play Measurement-Device-Independent Quantum Key Distribution}

\author{Yujun Choi}\email{cyj5595@gmail.com}
\affiliation{Center for Quantum Information, Korea Institute of Science and Technology (KIST), Seoul, 02792, Republic of Korea}
\affiliation{Department of Physics, Yonsei University, Seoul, 03722, Republic of Korea}

\author{Osung Kwon}
\affiliation{Center for Quantum Information, Korea Institute of Science and Technology (KIST), Seoul, 02792, Republic of Korea}

\author{Minki Woo}
\affiliation{Center for Quantum Information, Korea Institute of Science and Technology (KIST), Seoul, 02792, Republic of Korea}
\affiliation{Graduate School of IT Convergence, Ajou University, Suwon, 16499, Republic of Korea}

\author{Kyunghwan Oh}
\affiliation{Department of Physics, Yonsei University, Seoul, 03722, Republic of Korea}

\author{Sang-Wook Han}
\affiliation{Center for Quantum Information, Korea Institute of Science and Technology (KIST), Seoul, 02792, Republic of Korea}

\author{Yong-Su Kim}\email{yong-su.kim@kist.re.kr}
\affiliation{Center for Quantum Information, Korea Institute of Science and Technology (KIST), Seoul, 02792, Republic of Korea}

\author{Sung Moon}
\affiliation{Center for Quantum Information, Korea Institute of Science and Technology (KIST), Seoul, 02792, Republic of Korea}

\date{\today} 

\begin{abstract}
\noindent	
Quantum key distribution (QKD) guarantees unconditional communication security based on the laws of quantum physics. However, practical QKD suffers from a number of quantum hackings due to the device imperfections. From the security standpoint, measurement-device-independent quantum key distribution (MDI-QKD) is in the limelight since it eliminates all the possible loopholes in detection. Due to active control units for mode matching between the photons from remote parties, however, the implementation of MDI-QKD is highly impractical. In this paper, we propose a novel method to resolve the mode matching problem while minimizing the use of active control units. By introducing Plug-and-Play (P\&P) concept into MDI-QKD, the indistinguishability in spectral and polarization modes between photons can naturally be guaranteed. We show the feasibility of P\&P MDI-QKD with a proof-of-principle experiment.
\end{abstract}

\maketitle

\section{Introduction}

Quantum key distribution (QKD) has been focused as highly feasible technology in quantum information science since the first QKD protocol was introduced in 1984~\cite{bennett84}. QKD allows two remote parties (Alice and Bob) to generate secret keys with unconditional security by the laws of quantum physics. On the foundation of the immaculate theory, QKD has been developed in practical way and even made into full-packaged systems~\cite{idq}. However, doubts on the unconditional security of a QKD system have been continually raised. These doubts originate from gaps between ideal models and real devices of the QKD system. For example, an ideal QKD protocol requires ideal single-photon source and detectors, however, implementing them is impossible with current technology. These imperfections can be maliciously exploited, and make the QKD system vulnerable to various attacks~\cite{makarov06,lamas07,lydersen10,weier11,jain11,xu10,sun11,huang13,tang13,jouguet13}. In order to overcome such weaknesses, device independent quantum key distribution (DI-QKD) was proposed~\cite{mayers98,acin07}. The security of DI-QKD does not depend on the devices' characteristics, which means that any attack does not succeed even if the devices are not ideal, and thus DI-QKD always guarantees unconditional security. However, DI-QKD is highly impractical to implement since it is equivalent to the loophole-free Bell test which requires very high efficiency single-photon detection technology.

The recently proposed measurement-device-independent quantum key distribution (MDI-QKD) closes the practicality gap of DI-QKD while compromising some aspects of security~\cite{braunstein12,lo12}. While MDI-QKD can be vulnerable to the quantum hacking at the light sources, it closes all the possible loopholes in detection. Because single-photon detectors have been primary targets of the attacks, MDI-QKD significantly improves the security of the practical QKD system.

In the MDI-QKD protocol, Alice and Bob prepare singe-photon states independently, and send them to a third party (Charlie). The bit information is randomly encoded in one of the four states, $|0\rangle, |1\rangle,|+\rangle=\frac{1}{\sqrt{2}}(|0\rangle+|1\rangle),$ and $|-\rangle=\frac{1}{\sqrt{2}}(|0\rangle-|1\rangle)$. Then, Charlie performs Bell state measurements on the incoming photons, and publicly announces the measurement results to Alice and Bob. Finally, Alice and Bob can share secret keys after the classical post-processing such as error correction and privacy amplification. The security of MDI-QKD is based on the time-reversed entanglement-based QKD protocol, and thus, even if an eavesdropper (Eve) possesses Charlie, she cannot obtain any information about the secret keys~\cite{biham96,inamori02}.

There have been some experimental implementations of MDI-QKD both in laboratories~\cite{ferreira13,liu13,tang14,tang14_2} and on deployed fiber networks~\cite{rubennok13,tang14_3}. Very recently, the MDI-QKD with continuous variable also has been implemented~\cite{pirandola15}. The most crucial technique for implementing MDI-QKD is to make the photons sent by Alice and Bob indistinguishable~\cite{lo12}. In other words, the spectral, polarization, and temporal modes of the photons from Alice and Bob should be identical. Since the photons are prepared independently, however, it is difficult to make the photons identical. All the existing MDI-QKD experiments utilized active control units to match the modes~\cite{ferreira13,liu13,tang14,tang14_2,rubennok13,tang14_3}. For instance, to match the spectrum, frequency locked lasers with gas cells~\cite{tang14} or distributed feedback lasers with temperature controllers~\cite{ferreira13} were used. The polarization drift during the transmission and photon arrival timings are also continuously monitored and actively controlled~\cite{ferreira13,liu13,tang14,tang14_2,rubennok13,tang14_3}. Since the active control units are bulky and expensive to implement, it is necessary to solve the mode matching issue with a more convenient and inexpensive way.

In this article, we propose a new MDI-QKD scheme that solves the mode matching problem. By introducing the plug-and-play method~\cite{muller97} to the MDI-QKD, we can significantly reduce the use of active control units. Then, we will show the feasibility of P\&P MDI-QKD by demonstrating the proof-of-principle experiment that exploits a single laser for MDI-QKD.

\section{Plug-and-Play MDI-QKD}

\begin{figure}[t]
\centering
\includegraphics[width=3.4in]{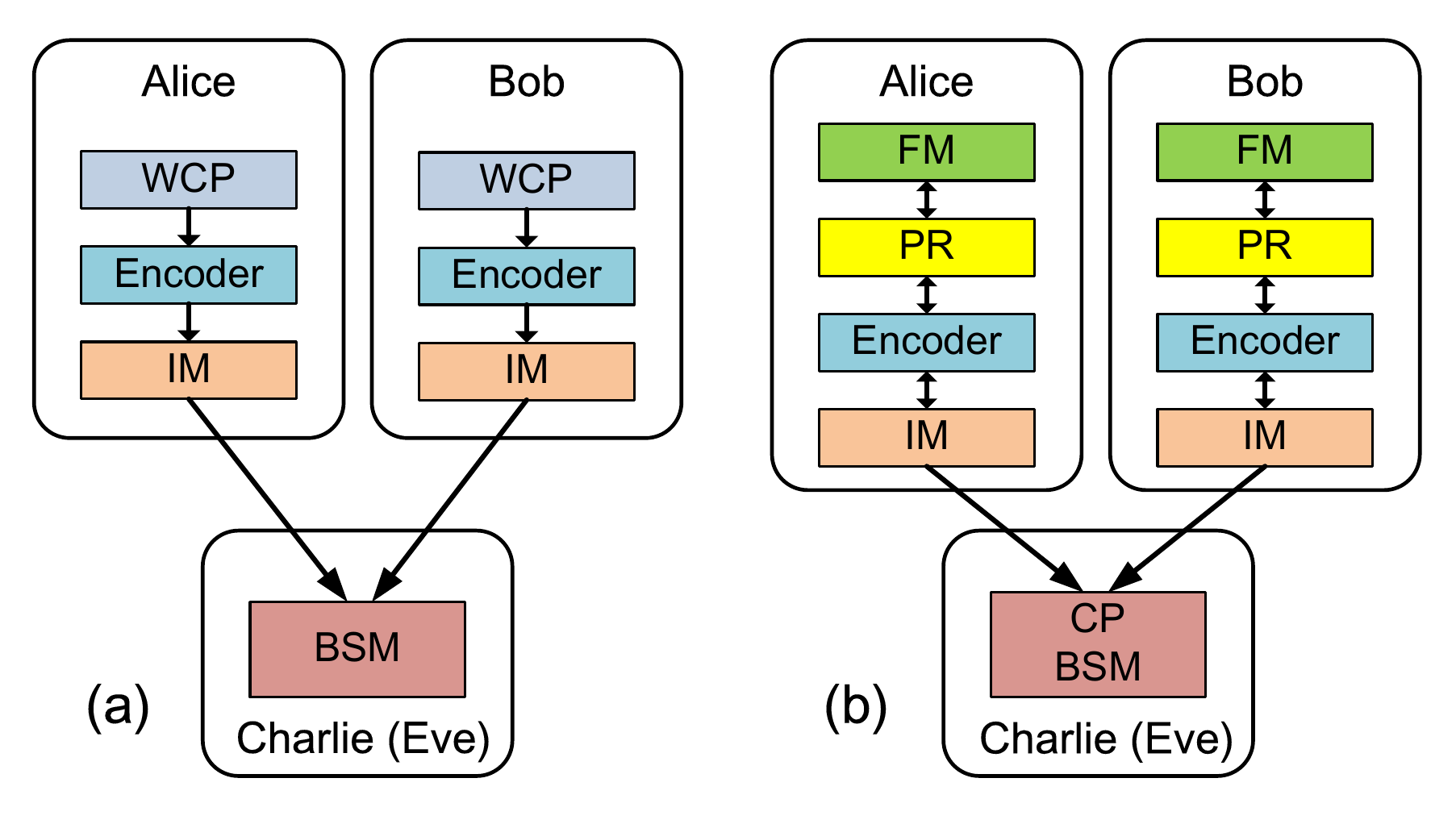}
\caption{Schematic diagrams of (a) conventional MDI-QKD protocol and (b) Plug-and-Play MDI-QKD protocol. WCP : weak  coherent pulse, CP : coherent pulse, Encoder : the device that encodes bit information, IM : intensity modulator, FM : faraday mirror, PR : phase randomizer, BSM : Bell state measurement. In a conventional MDI-QKD protocol, Alice and Bob prepare pulses independently, and send them to Charlie for Bell state measurement. In Plug-and-Play MDI-QKD, Charlie initially launches pulses to Alice and Bob, and then Alice and Bob reflect back the pulses to Charlie.
}\label{scheme}
\end{figure}

Figure~\ref{scheme}(a) shows the schematic of conventional MDI-QKD protocol with weak coherent pulses (WCP)~\cite{lo12}. An encoder randomly assigns one of the four BB84 states to the pulse. The four BB84 states can be either in the Z-basis ($|0\rangle$ and $|1\rangle$) or the X-basis ($|\pm\rangle=\frac{1}{\sqrt{2}}(|0\rangle\pm|1\rangle$). An intensity modulator (IM) generates decoy states~\cite{hwang03,lo05} of the pulses to protect the protocol from the photon number splitting attack. Alice and Bob transmit the pulses to Charlie, and then he performs Bell state measurements (BSM). Note that one can implement BSM that distinguishes two Bell states, $|\psi^+\rangle$ and $|\psi^-\rangle$, with linear optics and two-photon interference. This incomplete BSM is sufficient for implementing MDI-QKD. After performing the BSM, Charlie announces the results publicly, i.e., whether he got $|\psi^+\rangle=\frac{1}{\sqrt{2}}(|01\rangle+|10\rangle)$ or $|\psi^-\rangle=\frac{1}{\sqrt{2}}(|01\rangle-|10\rangle)$, and then Alice and Bob generate sifted keys only when their bases are identical.

%%%%%%%%%%%%%%%%%%%%%%%%%%%%%%%%%
%%%%%%%%%%%%%%%%%%%%%%%%%%%%%%%%%

It is interesting to note that there is an error rate of 25\% in X-basis caused by the pulses containing multi-photons of the coherent pulses~\cite{ferreira13}. Thus, with the WCP implementation, Alice and Bob generate secret keys only when they prepare the qubits in the Z-basis, and the X-basis is used to monitor Eve's attack. If one implements MDI-QKD with single-photon states, both the Z- and X-bases can be used for generating secret keys.

In the P\&P MDI-QKD scheme shown in Fig.~\ref{scheme}(b), the entire process is similar, but slightly different from that of the conventional MDI-QKD. First, Charlie sends strong coherent pulses to Alice and Bob through a 50:50 beamsplitter (not shown in Fig.~\ref{scheme}(b)). Each pulse is transmitted through an optical fiber, and then reflected by a Faraday mirror (FM) of Alice (Bob). The encoder and the IM randomly assign a bit and decoy states to the pulses. Note that the IM attenuates the optical pulses and sets the average photon numbers of the signal and decoy pulses to the proper values. Then, the pulse is sent back to Charlie through the same optical fiber. After that, BSM and post-processing are followed to obtain secret keys. It is notable that phase randomizers (PR) of Alice and Bob randomize the phase of the reflected pulse, which is necessary for decoy state protocol~\cite{lo05} and suppression of single-photon interference. Note that the suppression of single-photon interference is necessary to observe two-photon interference without distortion which is essential for MDI-QKD. The phase randomizers can be implemented with unsynchronized phase modulators~\cite{kim13,kim14}.

The P\&P MDI-QKD has remarkable features in the mode matching issue. Since the pulses of Alice and Bob are from the same laser, the spectral modes of the pulses are naturally identical. Moreover, the polarization drift during the optical fiber transmission can be automatically compensated due to the plug-and-play architecture~\cite{muller97}. One only needs to actively control the arrival timing of the pulses which is also essential for the conventional MDI-QKD. With these aspects, we can significantly reduce the efforts for mode matching with the suggested P\&P MDI-QKD.

It is necessary to discuss how the plug-and-play architecture affects the security of MDI-QKD. A rigorous security proof for P\&P MDI-QKD has been provided recently~\cite{xu15}. The only thing we must consider now is the security of P\&P MDI-QKD system being implemented with realistic devices. Since the plug-and-play architecture does not disturb the measurement setup of the MDI-QKD, unconditional security against any detector side channel attacks is guaranteed. On the other hand, the plug-and-play architecture may weaken the security against the attacks on the light sources, such as Trojan-horse and phase remapping attacks. These vulnerabilities can be circumvented by applying countermeasure techniques that are already developed for an ordinary plug-and-play QKD~\cite{xu10,gisin06,zhao08}. However, it is still an open question to develop robust countermeasures for plug-and-play QKD system~\cite{sajeed15}. We remark that the security issue for the P\&P MDI-QKD might be more complicated and interesting to study if Chalie is disguised by Eve and does not cooperate with Alice and Bob.

%Moreover, in preparation of inventing a new attack that emasculates already developed countermeasures, the security of realistic P\&P MDI-QKD system should be continuously researched like any other plug-and-play QKD systems~\cite{sajeed15}.

%%%%%%%%%%%%%%%%%%%%%%%%%%%%%%%%%
\begin{figure}[t]
\centering
\includegraphics[width=3.4in]{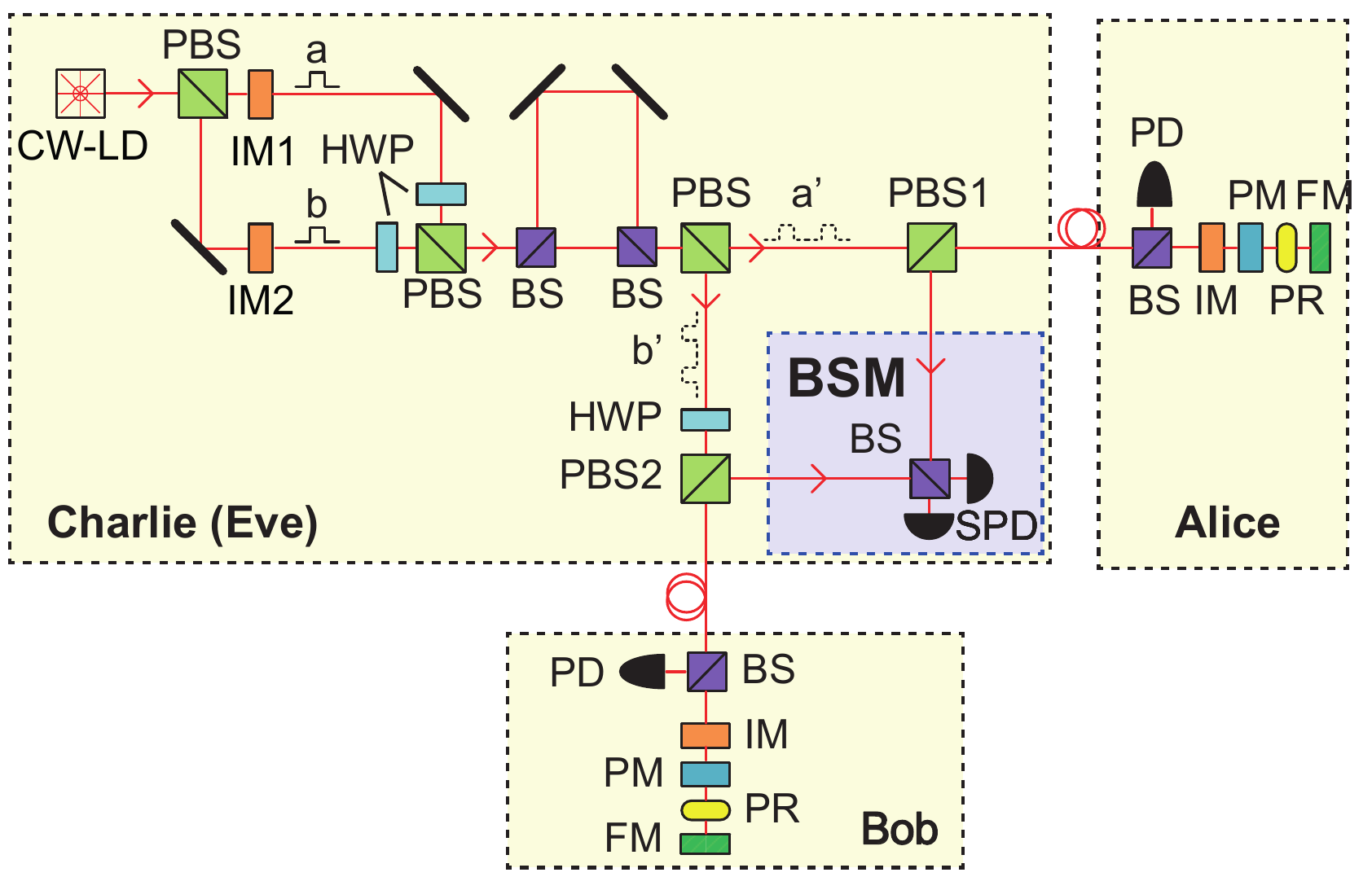}
\caption{A proposed setup for P\&P MDI-QKD using time-bin phase encoding. The synchronization for different distances between Charlie to Alice and Charlie to Bob is actively controlled with IM1 and IM2. CW-LD : continuous-wave laser diode, IM : intensity modulator, HWP : half wave plate, PBS : polarizing beamsplitter, BS : beamsplitter, PD : photodiode, PM : phase modulator, SPD : single-photon detector. Refer to the text for detailed explanation.
}\label{proposing}
\end{figure}
%%%%%%%%%%%%%%%%%%%%%%%%%%%%%%%%%

We propose a detailed implementation scheme for P\&P MDI-QKD using time-bin phase encoding~\cite{liu13,tang14_2,rubennok13,tang14_3} which is suitable for optical fiber quantum channels in Fig.~\ref{proposing}. Laser beam from a continuous-wave laser diode (CW-LD) goes through a Mach-Zehnder interferometer (MZI) which is made of two polarizing beamsplitters (PBSs) and half wave plates (HWPs). The MZI is used to generate two optical pulses for Alice and Bob, respectively. From a known initial polarization state of the laser beam, one can make two orthogonally polarized laser pulses at the output of the MZI. These orthogonal pulses are separated by a PBS later, and sent to Alice and Bob, respectively. Since one can operate two intensity modulators, IM1 and IM2, independently, it is possible to actively control the temporal delay between two orthogonally polarized pulses. In this way, the temporal mode matching at Charlie's BSM can be achieved and maintained. Note that the travel time of an optical pulse in an optical fiber is largely dependent on the temperature. For example, it has been shown that more than 30~ns of optical pulse travel time drift with less than 20~km of deployed optical fiber during QKD experiment~\cite{stucki11}. Therefore, the active control of the temporal delay is essential for the practical implementation of P\&P MDI-QKD.

The two orthogonally polarized pulses are sent to an asymmetric MZI (AMZI) in order to make the temporal superposition of the pulses, i.e., time-bins. The dotted two sequential pulses, a' and b', represent the time-bins. In contrast to the conventional time-bin phase MDI-QKD~\cite{liu13,tang14_2,rubennok13,tang14_3}, it is unnecessary to match the phase reference for X-basis between Alice and Bob because the time-bins are generated from a common AMZI. 

After the AMZI, a PBS separates two orthogonally polarized pulses and sends them to Alice and Bob, respectively. PBS1 and PBS2 are employed to clean up the polarization states of outgoing pulses from Charlie. Note that a HWP is employed before PBS2 in order to rotate the polarization of the pulses from vertical to horizontal. The intensity of the pulses sent to Alice and Bob is strong enough, thereby, a beam splitter (BS) and a photodiode (PD) can be used to detect the incoming pulses and make synchronization. The PD monitoring is also useful in preventing Trojan-horse attack~\cite{gisin06} and phase-remapping attack~\cite{xu10}. After the pulses are reflected at an FM, the polarization of the pulses changes orthogonally, and then, the phase of the pulses is randomized by a PR.

Before Alice (Bob) sends back the pulses to Charlie, she (he) encodes a bit into time-bin of pulses with a PM and IM. For X-basis, the PM modulates the relative phase between time-bins, either $0$ or $\pi$. One can implement Z-basis encoding by simply blocking either the fast or slow time-bins with an IM. Note that decoy state also can be implemented with an IM. The intensity of outgoing pulses should be properly reduced to guarantee the security of the protocol, which can be done with an IM as well.

The optical pulses sent back to Charlie are reflected at PBS1 and PBS2 thanks to the plug-and-play architecture~\cite{muller97}. Finally, Charlie performs Bell state measurements with two incoming optical pulses from Alice and Bob. Note that the Bell state measurements to the time-bin phase encoding can be achieved with a 50:50 BS and two single-photon detectors (SPDs)~\cite{ma12}. It is worth to remark that the timing between two pulses can be adjusted with IMs in the MZI.

\section{Experiment : a single laser MDI-QKD implementation}
\subsection{Experimental setup}

%%%%%%%%%%%%%%%%%%%%%%%%%%%%%%%%%
\begin{figure*}[t]
\includegraphics[width=6.5in]{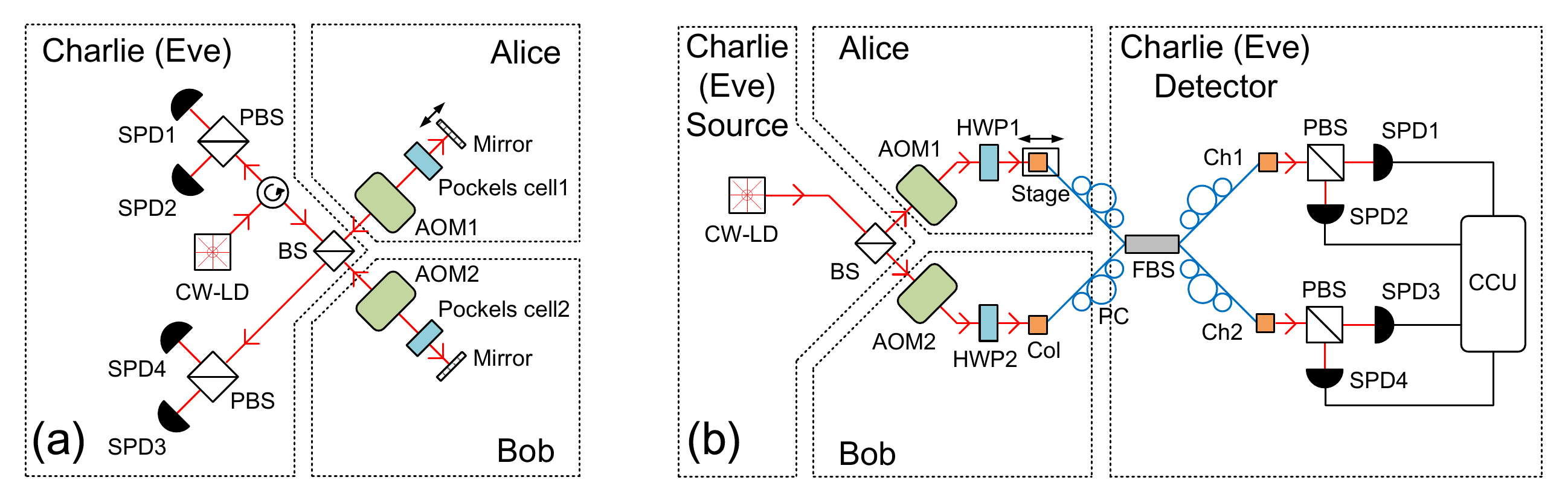}
\caption{(a) Experimental concept and (b) experimental setup of the proof-of-principle single laser MDI-QKD. CW-LD : continuous-wave laser diode, BS : beamsplitter, AOM : acousto-optic modulator, HWP : half waveplate, Col : collimating lens, Stage : translational stage, PC : polarization controller, FBS : fiber beamsplitter, Ch1,2 : Channel 1,2, PBS : polarizing beamsplitter, SPD : singe-photon detector, CCU : coincidence counting unit. Experimental concept is a physical implementation of P\&P MDI-QKD in bulk optics. Experimental setup is an extension of the experimental concept for expedience. All the experiment is done with the experimental setup.}\label{setup}
\end{figure*}
%%%%%%%%%%%%%%%%%%%%%%%%%%%%%%%%%

In this section, instead of proceeding an experiment with the proposed setup for P\&P MDI-QKD in Fig.~\ref{proposing}, we simply verify the feasibility of using a single laser for MDI-QKD mainly due to the lack of resources. The advantages of the P\&P MDI-QKD come from two aspects: First, the plug-and-play architecture provides the automatic polarization drift compensation and a common phase reference for time-bin phase encoding. Second, utilizing a common laser guarantees the identical spectral mode between optical pulses from Alice and Bob. The automatic polarization drift compensation has been shown through a lot of studies on plug-and-play QKD. On the other hand, using a single laser for MDI-QKD can be less familiar. Indeed there are only few studies about two-photon interference with a single laser that is essential for P\&P MDI-QKD implementation~\cite{kim13,kim14}. Although there is an experimental study showing the feasibility of Bell state measurement with time-bin phase encoding and fiber optical components~\cite{valivarthi14}, it lacks some important aspects that are essential for P\&P MDI-QKD, such as the relation to the original P\&P MDI-QKD concept, suppression of the first-order coherence, etc. Therefore, in this proof-of-principle experiment, we focus on showing the feasibility of using a single laser for MDI-QKD implementation with polarization encoding and bulk optical components in detailed process.

Figure~\ref{setup}(a) represents an experimental concept for a bulk-optic implementation of the MDI-QKD with polarization encoding. For the simplicity, we employ a multi-mode continuous-wave operating laser diode (CW-LD). Note that even though there is an ambiguity in temporal overlap of photons, it is known that two-photon interference which is essential for the MDI-QKD can be observed using a CW-LD~\cite{kim14}. The laser beam emitted from the CW-LD is transmitted toward a BS through an optical circulator indicated by a bending arrow in a circle. The beam is separated into two paths (Alice and Bob) by a BS. The separated beams go into acousto-optic modulators (AOM) which modulate the beams independently. The first-order diffracted beams go through high-speed Pockels cells that encode bits to the polarization modes of the beam. Since an AOM adds additional phase to the defracted beam relative to the driving RF signal, two independent AOMs suppress the single-photon interference \cite{kim13,kim14}. Note that, in the setup, the AOMs and the HWPs are substitutes for the PRs and the encoders in Fig.~\ref{scheme}(b), respectively. After encoding, the beams are reflected by mirrors, and sent back to Charlie. One can scan the position of the Alice's mirror in order to match the temporal modes. The BS, two PBSs, and four SPDs in Charlie are the components for BSM with polarization encoding. For BSM, $|\psi^+\rangle$ corresponds to the coincidence detection between SPD1 and SPD2 or between SPD3 and SPD4, and $|\psi^-\rangle$ corresponds to the coincidence detection between SPD1 and SPD4 or between SPD2 and SPD3~\cite{lo12}.

%%%%%%%%%%%%%%%%%%%%%%%%%%%%%%%%%
\begin{figure*}[t]
\includegraphics[width=6.5in]{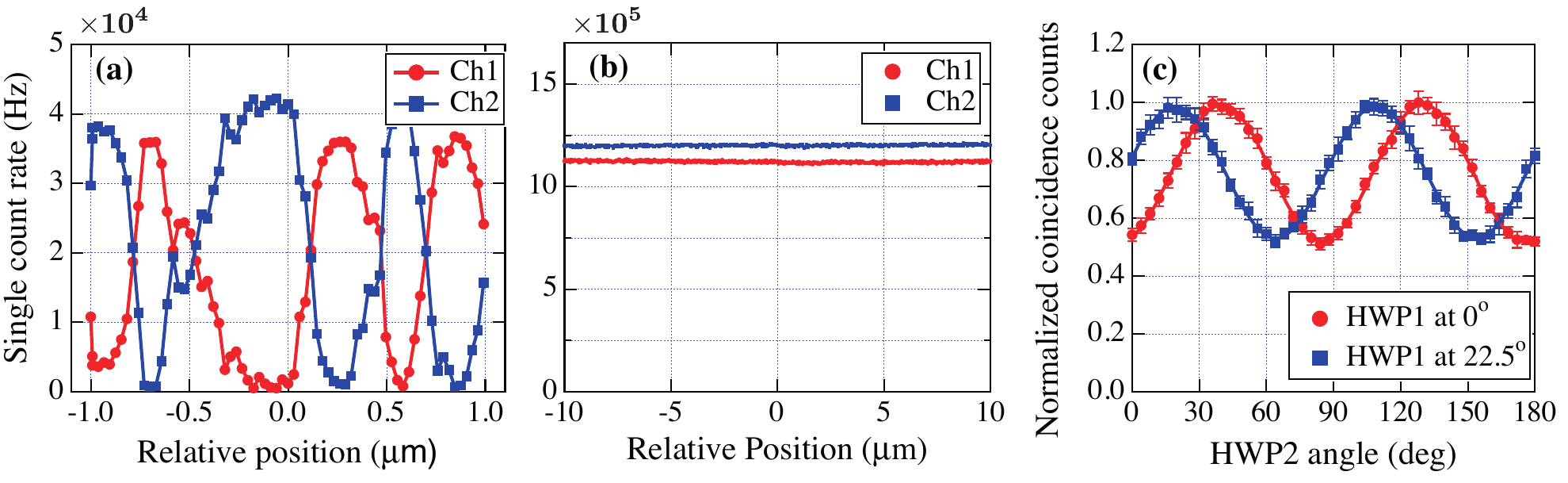}
\caption{Single-photon interference with (a) synchronized AOMs and (b) independently operating AOMs. The independently operating AOMs degrade single-photon interference. (c) Two-photon interference with independently operating AOMs. The two-photon interference is observed by rotating HWP2 while HWP1 is fixed at a certain angle. Solid curves and error bars denote fitted sine curves and standard deviations, respectively.}\label{interference}
\end{figure*}
%%%%%%%%%%%%%%%%%%%%%%%%%%%%%%%%%

In order to verify the feasibility of MDI-QKD using a single laser, we implement an experimental setup as shown in Fig.~\ref{setup}(b), which is a simple extension of the experimental concept of Fig.~\ref{setup}(a). Though the source and detector parts of Charlie are now spatially separated, the setup is equivalent to the concept. All the procedures are the same except that a fiber beamsplitter (FBS) is used for the BSM that makes the optical alignment easier. In Fig.~\ref{setup}(b), after passing through the HWPs, the beams interfere at the FBS. The optical path length can be adjusted by scanning Alice's collimating lens which is located before the FBS. Four polarization controllers (PC) are used to maintain the polarization mode in the fiber. The outputs of the FBS, Ch1 and Ch2, are transmitted to the PBSs and the SPDs for BSM. A coincidence counting unit (CCU) records single counts of each SPD's output and coincidence counts between different SPD's outputs~\cite{park15}.

In our experiment, we utilized a 780~nm CW-LD. Between CW-LD and the BS, a polarizer and neutral density filters were inserted to clean up the polarization mode and attenuate the laser beam so as to make it proper to the QKD experiment (not shown in Fig.~\ref{setup}(b)). Two AOMs were modulated by two AOM drivers operating independently. The operating RF frequencies of AOM drivers were identical (40~MHz) to minimize the frequency shift of the diffracted beams. By adjusting the RF power, the average photon numbers of Alice and Bob were set the same. Four SPDs are silicon avalanche photodiodes, which has quantum efficiency of about 50\% at 780~nm. The CCU was implemented with an FPGA, and the coincidence window of CCU is about 8~ns.

\subsection{Experimental result}

Two-photon interference is essential for the proposed scheme as for the conventional MDI-QKD because it is a fundamental element of the BSM. To make the optical path lengths of Alice and Bob equal, we adjust Alice's optical path length by scanning the collimating lens.

Since we are interested in the BSM with two-photon interference, single-photon interference should be washed out. This can be done by inserting independently modulated AOMs in the whole interferometer arms~\cite{kim13,kim14}. In order to measure the single-photon and two-photon interference, Ch1 and Ch2 were directly connected to two SPDs. When two AOMs are synchronized, one can see clear single-photon interference, see Fig.~\ref{interference}(a). As shown in Fig.~\ref{interference}(b), however, the single-photon interference is washed out when two AOMs are independently modulated.

After confirming the independently modulated AOMs efficiently suppress the single-photon interference, we observed two-photon interference. The two-photon interference was measured by changing one of the polarization states of two inputs while the other's polarization state is unchanged, see Fig.~\ref{interference}(c). The error bars and solid lines are the experimental standard deviations and sine fittings, respectively. The interference visibilities are measured to be $31.3\pm0.89\%$ and $30.1\pm0.77\%$ when HWP1 was fixed at $0\degree$ and $22.5\degree$, respectively. These visibilities come from the way we calculate for general interference fringe. When we compare these with the HOM dip, these correspond to $47.7\pm0.90\%$ and $46.3\pm0.78\%$, and are close to the classical limit of two-photon interference visibility, $50\%$.

%%%%%%%%%%%%%%%%%%%%%%%%%%%%%%%%%
\begin{figure}[b]
\centering
\includegraphics[width=3.2in]{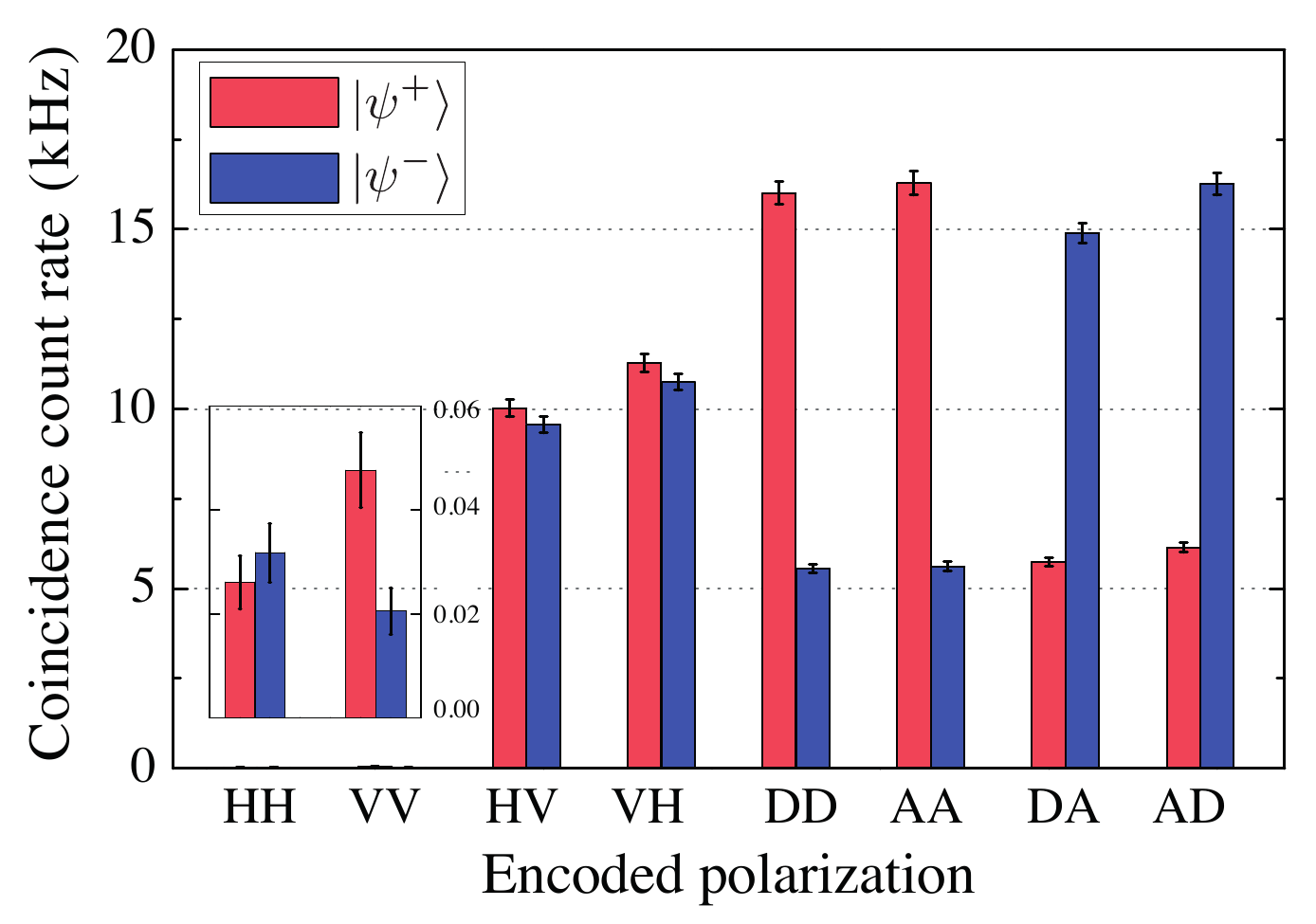}
\caption{Results of Bell state measurement when average photon number $\mu$ is 0.5. The average photon numbers of Alice and Bob are equal. The horizontal axis denotes the polarization encoding by Alice and Bob, e.g., HV means that Alice and Bob send horizontal and vertical polarization, respectively. The inset represents zoomed-in coincidence count rates for HH and VV cases. The unit of vertical axis of inset is same as that of main graph (kHz). Solid bars and error bars denote average coincidence count rates and standard deviations, respectively.}\label{MDI}
\end{figure}
%%%%%%%%%%%%%%%%%%%%%%%%%%%%%%%%%

Finally, we performed BSM and estimated expected quantum bit error rate (QBER) according to the MDI-QKD protocol.The BSM setup shown in Fig.~\ref{setup}(b) was restored for this purpose. During the experiment, the average photon numbers of Alice and Bob are set to $\mu=0.5$. The BSM result is shown in Fig.~\ref{MDI}. The horizontal axis represents the encoded polarization states by Alice and Bob. For example, `HV' is that Alice and Bob send horizontal and vertical polarization states, respectively. For each polarization states, the coincidence counts were accumulated for 10 minutes. From the data, we obtained average coincidence count rates and standard deviations, which are indicated by solid bars and error bars of Fig.~\ref{MDI}, respectively. The inset is zoomed-in coincidence count rates of HH and VV cases whose coincidence count rates are less than 60~Hz, which means that they are negligibly small compared to that of other cases (HV,VH,...,AD).

It is necessary to estimate the expected QBER from the data in order to show the feasibility of the single laser MDI-QKD. In the Z-basis, the erroneous coincidence counts are both $|\psi^{+}\rangle$ and $|\psi^{-}\rangle$ for HH or VV. In the X-basis, erroneous coincidence counts are $|\psi^{-}\rangle$ for DD or AA and $|\psi^{+}\rangle$ for DA or AD. Based on that, we can calculate the QBER for each case. The QBERs are given by
%%%%%%%%%%%%%%%%%%%%%%%%%%%%%%%%%%%%
\begin{equation}
E_Z = \frac{C_{HH} + C_{VV}}{C_{HH} + C_{VV} + C_{HV} + C_{VH}}
\end{equation}
\begin{equation}
E_X = \frac{C_{DD}^{-} + C_{AA}^{-} + C_{DA}^{+} +
C_{AD}^{+}}{C_{DD} + C_{AA} + C_{DA} + C_{AD}}
\end{equation}
%%%%%%%%%%%%%%%%%%%%%%%%%%%%%%%%%%%%%
where $E_Z$ and $E_X$ stands for the QBER of the Z- and X-basis, respectively. $C_{ij} = C_{ij}^{+} + C_{ij}^{-}$ where $C_{ij}^{\pm}$ is the coincidence count corresponding to $|\psi^{\pm}\rangle$, and the subscript \textit{ij} is the encoding polarization states by Alice and Bob, respectively. The experimental QBERs for $\mu=0.3$ and 0.5 as well as the theoretical one are shown in Table~\ref{qber}. The QKD implementation with different average photon numbers corresponds to the decoy states implementation which enhances the security against photon number splitting attack~\cite{hwang03,lo05}. Noting the theoretical QBERs, 0\% for Z-basis and 25\% for X-basis, the experimental QBERs are small enough to verify the feasibility of the single laser MDI-QKD.

%%%%%%%%%%%%%%%%%%%%%%%%%%%%%%%%%
\begin{table}[t]
\caption{Experimental and theoretical QBERs of single laser MDI-QKD}
\begin{center}
\begin{tabularx}{3.4in}{@{}*4{>{\centering\arraybackslash}X}@{}}
\hline \hline
& \multicolumn{3}{c}{QBER} \\
\cline{2-4}
Basis  & $\mu$=0.3  & $\mu$=0.5 & Theory \\
\hline
Z  & 0.37$\pm$0.05\%  & 0.30$\pm$0.03\%  & 0    \\
X  & 26.5$\pm$0.46\%  & 26.7$\pm$0.35\%   & 25\%  \\
\hline \hline
\end{tabularx}
\end{center}
\label{qber}
\end{table}
%%%%%%%%%%%%%%%%%%%%%%%%%%%%%%%%%

\section{Conclusion}

The P\&P MDI-QKD has several advantages in comparison to the conventional MDI-QKD. The most important advantage is that one can significantly reduce the use of active control units. In time-bin phase encoding, for example, the spectral mode matching is naturally achieved by utilizing the same laser. The phase reference of time-bins pulses sent by Alice and Bob are also automatically matched since Charlie can possess the interferometer. These advantages will be a great merit for the practicality. Moreover, the plug-and-play architecture compensates the polarization drift during the transmission. This robustness will provide superior performance in implementing the MDI-QKD on deployed optical fibers. Finally, P\&P MDI-QKD is appropriate for realizing MDI-QKD network. Since Charlie (server) has a source and detectors, Alice and Bob (users) need only encoding devices. If the server is located at a control station, users can communicate securely with inexpensive and compact size devices.

In this paper, we proposed a complete implementation setup including all the spectral, polarization, and temporal mode matching methods. Moreover, our proof-of-principle experiment clearly shows the feasibility of using a single laser for implementing P\&P MDI-QKD. With the rigours security proof of P\&P MDI-QKD~\cite{xu15} and our results, we believe that P\&P MDI-QKD will facilitate practical implementation of MDI-QKD, and contribute to the ultimate secure communication.

\section{Acknowledgement}
This work was supported by the ICT R\&D programs of MSIP/IITP$[$10044559$, $2014-044-014-002$]$, and the KIST Research Programs (2E26410, 2V04600, 2N41850).

\end{document}